\documentclass[conference]{IEEEtran}
\IEEEoverridecommandlockouts

\usepackage{cite}
\usepackage{amsmath,amssymb,amsfonts}
\usepackage{subcaption}
\usepackage{algorithmic}
\usepackage{graphicx}
\usepackage{textcomp}
\usepackage{xcolor}
\usepackage{float}
\usepackage{graphicx}
\usepackage{subcaption}
\usepackage{url}
\usepackage{multirow}
\usepackage{soul,color}
\usepackage{amsmath}
\usepackage{subfig}
\usepackage{soul}
\usepackage{xcolor}
\usepackage{booktabs}
\def\BibTeX{{\rm B\kern-.05em{\sc i\kern-.025em b}\kern-.08em
    T\kern-.1667em\lower.7ex\hbox{E}\kern-.125emX}}

\usepackage{float}
\begin{document}

\title{An Attention-Based Framework for Alzheimer’s Disease Classification Using Resting-State fMRI\\
\thanks{Harshiddhi Pathak, Manjunatha Mahadevappa are with School of Medical Science and Technology, and Mrinal Acharya with Dr B C Roy Multi Speciality Medical Research Centre, Indian Institute of Technology Kharagpur, India. 
Gowtham is affiliated with the Department of Computer Science and Engineering at the Indian Institute of Technology Gandhinagar..
(e-mail:harshiddhipathak@gmail.com, gowtham12.reddy@gmail.com, 
mrinal@bcrmrc.iitkgp.ac.in,
mmaha2@smst.iitkgp.ac.in).\\
**First two authors contributed equally}
}
\author{
    \IEEEauthorblockN{
        Harshiddhi Pathak, Graduate Student Member, IEEE,
        Gowtham Reddy N, Member, IEEE, \\
        Mrinal Acharya,
        Manjunatha Mahadevappa, Senior Member, IEEE
    }
}

\maketitle

\begin{abstract}
Accurate identification of Alzheimer’s disease (AD) using resting-state functional magnetic resonance imaging (rs-fMRI) remains challenging due to the high dimensionality, noise, and complex inter-regional dependencies inherent in functional brain connectivity, which limit the effectiveness of traditional approaches based on handcrafted connectivity features or conventional machine learning models. In this work, we present an attention-based deep learning framework for Alzheimer’s disease classification that operates directly on rs-fMRI functional connectivity matrices by treating brain regions as tokens and employing a Transformer-inspired self-attention mechanism to model long-range and global functional dependencies across distributed brain networks. The proposed framework learns discriminative functional representations without reliance on manual feature engineering and is evaluated on a longitudinal cohort from the Alzheimer’s Disease Neuroimaging Initiative (ADNI) comprising cognitively normal and Alzheimer’s disease subjects with multiple visits. A subject-wise evaluation protocol is adopted to prevent information leakage across visits, and class-weighted optimization is incorporated to address mild class imbalance. Experimental results for binary AD versus cognitively normal classification demonstrate that the proposed attention-based rs-fMRI model achieves an accuracy of 88.95\% and a ROC-AUC of 0.90, along with a favorable precision–recall balance, highlighting the effectiveness of self-attention–driven functional connectivity modeling as a robust and interpretable approach for Alzheimer’s disease detection using resting-state fMRI.
\end{abstract}


\section{Introduction}\label{sec1}

\IEEEPARstart{A}{lzheimer’s} disease (AD) is a progressive neurodegenerative disorder and one of the most common causes of dementia worldwide, leading to severe cognitive decline and substantial socioeconomic burden. Early and reliable identification of AD remains a major clinical challenge, as disease-related brain alterations emerge gradually and vary across individuals. Neuroimaging has therefore become a cornerstone for computer-aided diagnosis, enabling non-invasive characterization of both structural degeneration and functional disruption in the brain.

Large-scale initiatives such as the Alzheimer’s Disease Neuroimaging Initiative (ADNI) have played a pivotal role in advancing Alzheimer’s research by providing standardized, longitudinal magnetic resonance imaging (MRI) and functional MRI (fMRI) data across multiple disease stages \cite{jack2008adni,adnioverview2008}. Early neuroimaging studies predominantly focused on handcrafted structural MRI features such as regional brain volumes and cortical thickness, revealing consistent atrophy patterns in Alzheimer’s disease. While informative, such approaches rely on predefined representations and are limited in capturing complex inter-regional interactions.

The growing availability of resting-state functional MRI (rs-fMRI) has enabled the investigation of functional connectivity alterations associated with Alzheimer’s disease. Traditional pipelines typically employed correlation-based connectivity measures followed by classical machine learning classifiers, revealing disrupted large-scale brain networks in AD \cite{szekely2009distance}. However, these methods are sensitive to noise, depend heavily on handcrafted connectivity summaries, and struggle to model the high-dimensional and relational nature of functional brain networks. More recent deep learning approaches have addressed some of these limitations by learning functional connectivity representations directly from rs-fMRI data, demonstrating improved classification performance \cite{zhang2021deepfmri,diagnostics2023ad}. Nevertheless, many of these studies remain limited  and do not exploit complementary anatomical information.


Attention-based architectures, particularly those inspired by the Transformer framework \cite{vaswani2017attention}, have recently gained increasing attention in medical imaging due to their ability to model long-range dependencies and learn adaptive, context-aware representations. In the context of resting-state fMRI, self-attention mechanisms are especially well suited for capturing complex and distributed functional interactions among brain regions, enabling the model to selectively emphasize disease-relevant connectivity patterns associated with neurodegeneration.
Despite their potential, many existing rs-fMRI–based Alzheimer’s disease studies rely on handcrafted connectivity summaries or conventional deep learning architectures that struggle to model global inter-regional dependencies. Moreover, attention-based approaches applied to functional neuroimaging often lack longitudinal evaluation protocols and do not explicitly address class imbalance, both of which are critical for clinically meaningful and robust disease classification.

Motivated by these limitations, this work proposes an attention-based deep learning framework for Alzheimer’s disease classification using resting-state fMRI functional connectivity. The proposed model treats brain regions as tokens and employs self-attention to learn global functional representations directly from connectivity matrices. Using longitudinal data from the ADNI cohort and a class-weighted optimization strategy, the framework aims to provide a robust, interpretable, and scalable solution for functional connectivity-driven Alzheimer’s disease detection.

\subsection*{Key Contributions}

The key contributions of this work are summarized as follows:
\begin{itemize}
    \item An attention-based deep learning framework is presented for Alzheimer’s disease classification that operates directly on resting-state fMRI functional connectivity matrices, modeling brain regions as tokens to capture global inter-regional dependencies.
    
    \item A self-attention mechanism is employed to learn long-range functional interactions across distributed brain networks, enabling adaptive identification of disease-relevant connectivity patterns without reliance on handcrafted features.
    
    \item A subject-wise longitudinal evaluation protocol combined with class-weighted optimization is adopted to ensure realistic generalization assessment and robust learning under mild class imbalance, reflecting clinically relevant data distributions.
    
    \item Experimental validation on the ADNI cohort demonstrates strong and balanced classification performance, highlighting the effectiveness of attention-driven functional connectivity modeling for rs-fMRI–based Alzheimer’s disease detection.
\end{itemize}

\section{Literature Review}\label{Literature}
Early approaches to rs-fMRI-based AD classification relied on handcrafted  functional connectivity features derived from pairwise ROI correlations, combined with conventional classifiers such as support vector machines, which while revealing disrupted large-scale networks including the default mode network were limited in their ability to model non-linear inter-regional dependencies~\cite{dennis2014functional}. Deep learning methods subsequently addressed some of these limitations: Zhang et al.~\cite{zhang2021deepfmri} demonstrated hierarchical representation learning from rs-fMRI for multi-stage AD classification, while Ahmed et al.~\cite{diagnostics2023ad} applied convolutional architectures to functional connectivity matrices, and Li et al.~\cite{LI2020280} leveraged 4D fMRI temporal dynamics within a deep network. However, convolutional architectures impose local receptive field constraints that are ill-suited to the globally distributed nature of functional connectivity disruptions in AD.

Graph neural networks (GNNs) have been proposed to explicitly encode the topology of functional brain networks by treating ROIs as nodes and 
connectivity strengths as edge weights~\cite{chang2021spectral}, enabling message passing across anatomically or functionally linked regions. Despite their promise, GNN-based methods require a predefined adjacency matrix constructed via thresholding or anatomical priors, which introduce snsitivity to graph construction choices. Transformer-based architectures~\cite{vaswani2017attention}, with their permutation-invariant self-attention mechanism, offer an alternative that learns global inter-regional interactions directly from data without imposing structural priors. Recent work has demonstrated the effectiveness of 
Transformer encoders for functional connectivity classification ~\cite{kan2022brain}, motivating the attention-based formulation adopted in this work.

\section{Dataset Outline}\label{dataset}

This study employs neuroimaging data obtained from the ADNI, a large, publicly available, longitudinal database established to support the investigation of Alzheimer’s disease progression and normal aging \cite{jack2008adni,adnioverview2008}. Data from the ADNI-GO, ADNI-2, and ADNI-3 phases were incorporated to ensure consistency in acquisition protocols and longitudinal coverage across subjects.\footnote{ADNI database: \url{https://adni.loni.usc.edu/}}

The final cohort consisted of 60 participants, including 30 subjects diagnosed with Alzheimer’s disease (AD) and 30 cognitively normal (CN) control subjects. Each participant contributed exactly two clinical visits acquired at an interval of approximately one year, resulting in a total of 120 visit-level samples. Each visit contained both a structural MRI scan and a resting-state fMRI scan, yielding a total of 120 T1-weighted MRI images and 120 rs-fMRI acquisitions for analysis. This controlled longitudinal design enables the investigation of disease-related patterns while maintaining balanced temporal sampling across both groups. The mean age of the AD cohort was $73.5 \pm 7.5$ years, while the cognitively normal cohort had a mean age of $72.8 \pm 6.9$ years, thereby minimizing potential confounding effects related to normal aging.

For each subject and visit, two complementary neuroimaging modalities were analyzed. Structural brain information was obtained using T1-weighted MRI acquired with magnetization prepared rapid gradient echo (MPRAGE) sequences, following standardized ADNI acquisition protocols. These high-resolution anatomical images provide detailed characterization of cortical and subcortical structures and were used for the extraction of structural biomarkers associated with neurodegenerative changes in Alzheimer’s disease. In addition, resting-state functional MRI (rs-fMRI) data were collected under eyes-open resting conditions to capture spontaneous low-frequency blood-oxygen-level-dependent (BOLD) signal fluctuations, enabling the analysis of functional connectivity patterns across distributed brain networks.

All imaging data were originally provided in Digital Imaging and Communications in Medicine (DICOM) format and were converted to Neuroimaging Informatics Technology Initiative (NIfTI) format prior to analysis to ensure compatibility with standard neuroimaging software. The dataset was organized in a subject-wise and visit-wise hierarchical structure, enabling accurate alignment between rs-fMRI data and the corresponding T1-weighted MRI scans for preprocessing purposes, while supporting subject-wise longitudinal evaluation in subsequent modeling stages.

\section{Method}\label{method}
\raggedbottom

\subsection{rs-fMRI Preprocessing and Functional Connectivity Extraction}
\label{sec:fmri_preprocessing}
\begin{figure}[t]
    \centering
    \includegraphics[width=1.0\linewidth]{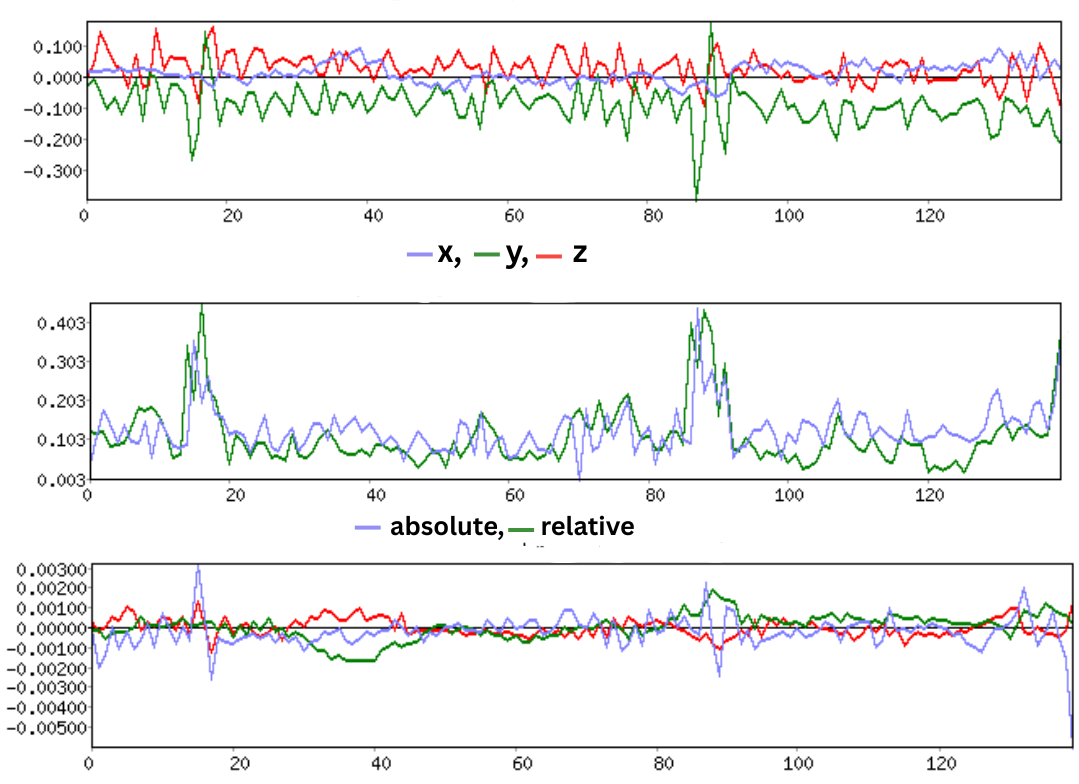}
    \caption{Head motion correction assessment for a representative Alzheimer’s disease (AD) subject obtained using FSL MCFLIRT. The top panel shows estimated translational head motion along the $x$, $y$, and $z$ axes (in millimeters), the middle panel depicts mean absolute and relative displacement between successive volumes, and the bottom panel presents estimated rotational motion around the three axes (in radians) across time. The overall low displacement values indicate effective motion correction and acceptable head motion during rs-fMRI acquisition.}
    \label{fig:motion_correction}
\end{figure}

Resting-state fMRI preprocessing was performed using FSL, following established neuroimaging practices. The pipeline began with slice timing correction to compensate for temporal acquisition offsets between slices, followed by motion correction using MCFLIRT, in which all volumes were aligned to a reference volume to mitigate the effects of head motion. The estimated translational and rotational motion parameters, along with mean absolute and relative displacement, were visually inspected to assess motion severity and preprocessing quality, ensuring that head motion remained within acceptable limits for reliable functional connectivity analysis (Figure~\ref{fig:motion_correction}). The same preprocessing pipeline was consistently applied to both Alzheimer’s disease (AD) and cognitively normal (CN) subjects, and the figure illustrates representative motion characteristics for an AD subject.

\begin{figure}[t]
\centering

\begin{subfigure}{0.48\textwidth}
    \centering
    \includegraphics[width=\linewidth]{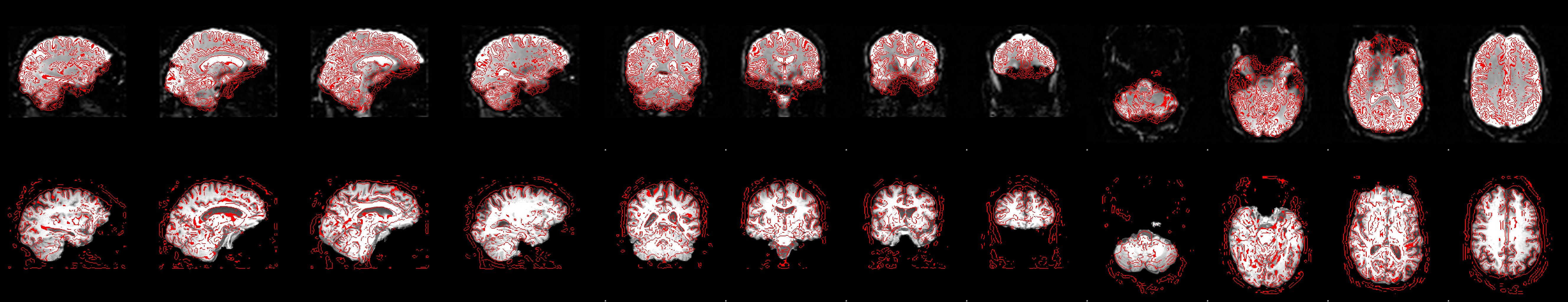}
    \caption{rs-fMRI to T1-weighted MRI}
\end{subfigure}
\hfill
\begin{subfigure}{0.48\textwidth}
    \centering
    \includegraphics[width=\linewidth]{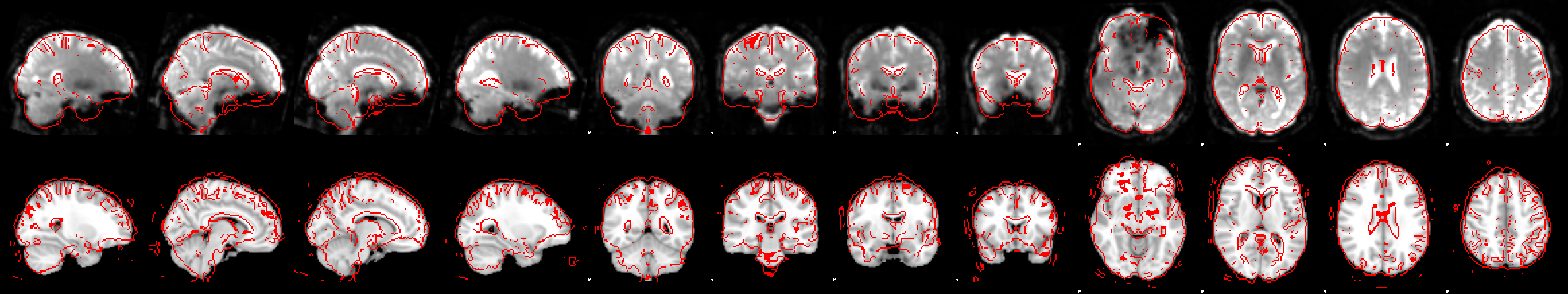}
    \caption{rs-fMRI to MNI152 space}
\end{subfigure}

\vspace{0.3cm}

\begin{subfigure}{0.48\textwidth}
    \centering
    \includegraphics[width=\linewidth]{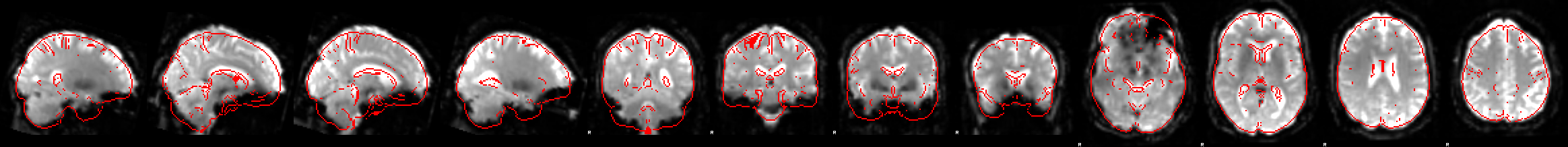}
    \caption{rs-fMRI to MNI152 (alternate slices)}
\end{subfigure}
\hfill
\begin{subfigure}{0.48\textwidth}
    \centering
    \includegraphics[width=\linewidth]{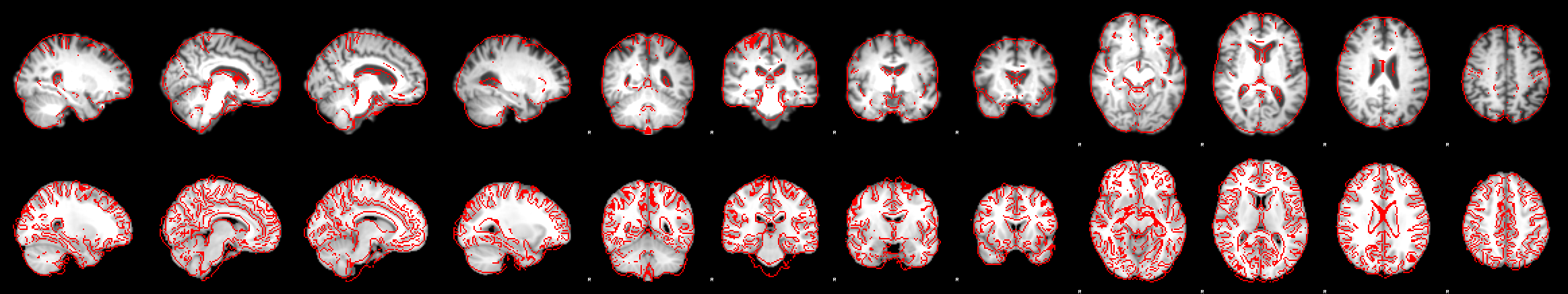}
    \caption{T1-weighted MRI to MNI152 space}
\end{subfigure}

\caption{Spatial registration results for a representative Alzheimer’s disease (AD) subject. The figure illustrates alignment of rs-fMRI data to the subject’s high-resolution T1-weighted MRI and subsequent normalization to the MNI152 standard space using FSL FLIRT and FNIRT. Red contours indicate accurate anatomical correspondence across modalities.}
\label{fig:registration}

\end{figure}

Each subject’s rs-fMRI data were subsequently coregistered to the corresponding high-resolution T1-weighted MRI to ensure accurate anatomical correspondence between functional and structural images. The functional images were then spatially normalized to the MNI152 standard space using a combination of linear (FLIRT) and nonlinear (FNIRT) registration.
Representative registration results for an AD subject are shown in Figure~\ref{fig:registration}, illustrating accurate alignment across functional, structural, and standard spaces.

After normalization, spatial smoothing was applied using a Gaussian kernel with a full-width at half-maximum (FWHM) of 4--5~mm to improve signal-to-noise ratio and reduce residual inter-subject variability. Temporal band-pass filtering was then performed in the frequency range of 0.01--0.1~Hz to retain low-frequency BOLD signal fluctuations associated with intrinsic functional connectivity.

\begin{figure}[t]
    \centering
    \includegraphics[width=0.7\linewidth]{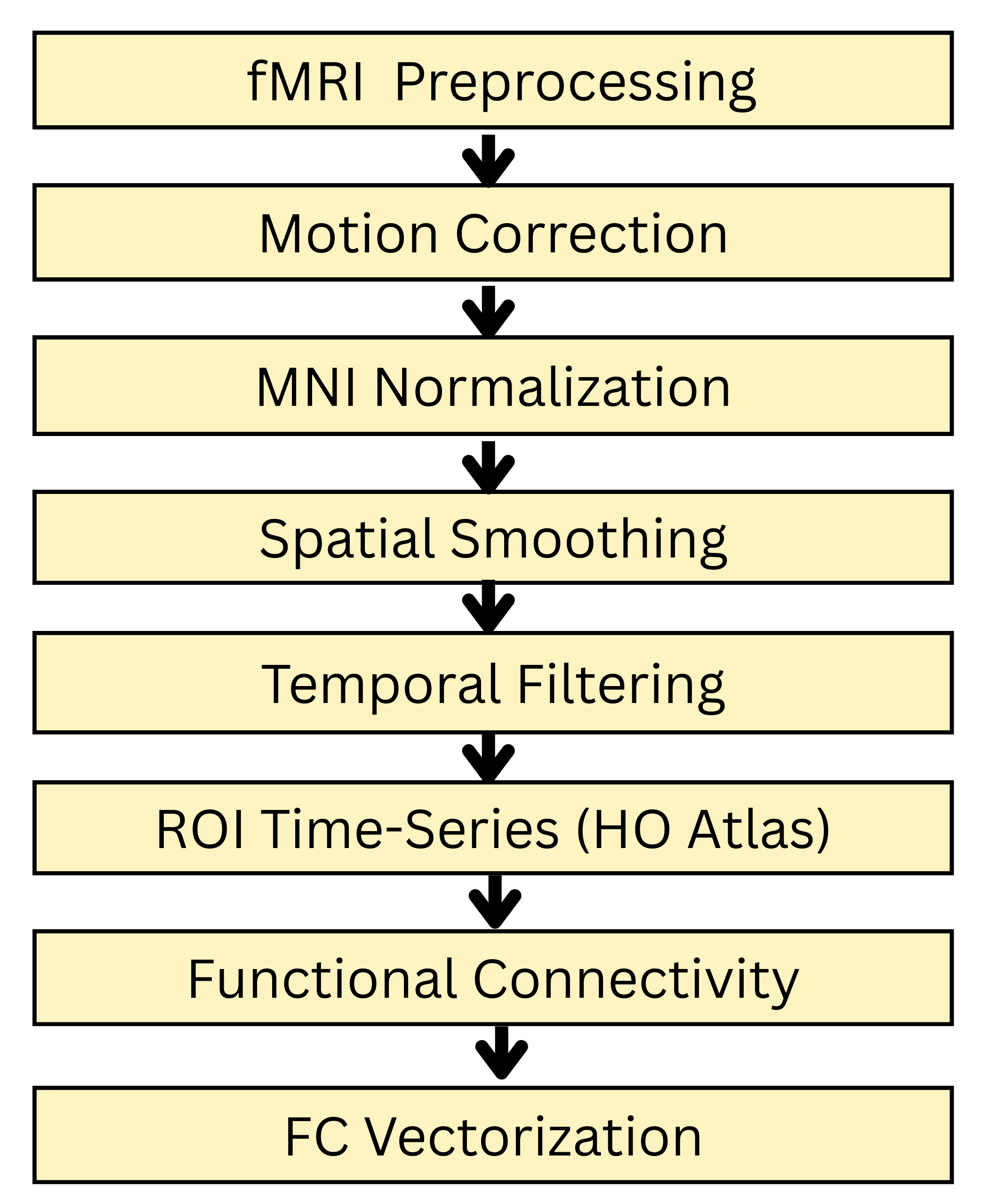}
    \caption{Overview of the resting-state fMRI preprocessing and functional connectivity extraction pipeline, including motion correction, spatial normalization, temporal filtering, atlas-based parcellation, and functional connectivity computation used for attention-based modeling.}
    \label{fig:fmri_preprocessing}
\end{figure}

The preprocessed rs-fMRI data were parcellated into 69 cortical and subcortical regions using the Harvard--Oxford atlas provided by FSL. Mean BOLD time series were extracted for each ROI, and functional connectivity matrices were computed using pairwise Pearson correlation coefficients. ROIs exhibiting zero variance were excluded, and missing connections were handled by zero-filling to maintain consistent dimensionality across subjects.
An overview of the complete rs-fMRI preprocessing and functional connectivity extraction workflow is illustrated in Figure~\ref{fig:fmri_preprocessing}.

\subsection{Attention-based fMRI classification Model}\label{Attention-based fMRI classification Model}

This section describes the proposed attention-based deep learning framework for Alzheimer’s disease classification using resting-state fMRI data. The model operates directly on functional connectivity representations derived from rs-fMRI and leverages self-attention mechanisms to capture long-range dependencies among distributed brain regions. The overall architecture consists of three main components: (i) functional connectivity feature representation, (ii) a self-attention module for modeling inter-regional interactions, and (iii) a classification head optimized using a class-weighted objective.
\subsection{Feature Representation of Functional Connectivity}
\label{sec:fmri_feature_representation}

Resting-state fMRI data were represented using region-wise functional connectivity (FC) matrices computed from preprocessed BOLD time series. For each subject visit, pairwise Pearson correlation coefficients were estimated between predefined regions of interest (ROIs), resulting in a subject-specific FC matrix of size $N \times N$, where $N$ denotes the total number of ROIs.

Due to occasional signal dropout or preprocessing failures in certain regions, a subset of ROIs may be invalid for a given visit. To ensure a consistent input dimensionality across all samples, only valid ROI pairs were retained during FC computation. The resulting reduced FC matrix was then embedded back into a full $N \times N$ matrix by zero-filling entries corresponding to invalid ROI indices. 

Formally, each visit-level input sample is represented as
\begin{equation}
\mathbf{X}_{\text{fMRI}} \in \mathbb{R}^{N \times N},
\end{equation}
where $\mathbf{X}_{\text{fMRI}}$ denotes the reconstructed functional connectivity matrix. This matrix-based representation allows the model to explicitly capture inter-regional functional interactions and serves as the input to the subsequent attention-based feature learning module.
Each sample is associated with a binary diagnostic label $y \in \{0,1\}$, where $y=1$ denotes Alzheimer’s disease (AD) and $y=0$ denotes cognitively normal (CN).
\subsection{Self-Attention Mechanism for rs-fMRI Features}
\label{sec:fmri_self_attention}

Given a reconstructed functional connectivity matrix
$\mathbf{X}_{\text{fMRI}} \in \mathbb{R}^{N \times N}$,
each row of the matrix corresponds to the functional connectivity profile of a single ROI with respect to all other regions. In the proposed framework, each ROI is treated as an individual token, enabling attention-based modeling of inter-regional dependencies.

To project each ROI connectivity profile into a latent embedding space, a learnable linear transformation is applied independently to each row of the FC matrix:
\begin{equation}
\mathbf{z}_i = \phi\left(\mathbf{W}_e \mathbf{x}_i + \mathbf{b}_e\right),
\end{equation}
where $\mathbf{x}_i \in \mathbb{R}^{N}$ denotes the $i$-th row of the FC matrix, $\mathbf{W}_e \in \mathbb{R}^{d \times N}$ is a trainable projection matrix, $\mathbf{b}_e$ is a bias term, $d$ is the embedding dimension, and $\phi(\cdot)$ denotes a nonlinear activation function. This operation results in a sequence of ROI embeddings
\begin{equation}
\mathbf{Z} = [\mathbf{z}_1, \mathbf{z}_2, \dots, \mathbf{z}_N] \in \mathbb{R}^{N \times d}.
\end{equation}

To model global functional interactions among ROIs, a multi-head self-attention mechanism is applied to the embedded token sequence. Query, key, and value matrices are computed as:
\begin{equation}
\mathbf{Q} = \mathbf{Z}\mathbf{W}_Q, \quad
\mathbf{K} = \mathbf{Z}\mathbf{W}_K, \quad
\mathbf{V} = \mathbf{Z}\mathbf{W}_V,
\end{equation}
where $\mathbf{W}_Q, \mathbf{W}_K, \mathbf{W}_V \in \mathbb{R}^{d \times d}$ are learnable projection matrices.

The self-attention operation is defined as:
\begin{equation}
\text{Attention}(\mathbf{Q}, \mathbf{K}, \mathbf{V}) =
\text{softmax}\!\left( \frac{\mathbf{Q}\mathbf{K}^{\top}}{\sqrt{d}} \right)\mathbf{V}.
\end{equation}

This formulation allows each ROI to dynamically attend to other regions based on functional relevance, enabling the model to capture long-range dependencies and network-level interactions that are known to be disrupted in Alzheimer’s disease. Multi-head attention is employed to allow the model to jointly attend to information from different representational subspaces.

\begin{figure}[t]
    \centering
    \includegraphics[width=0.98\linewidth]{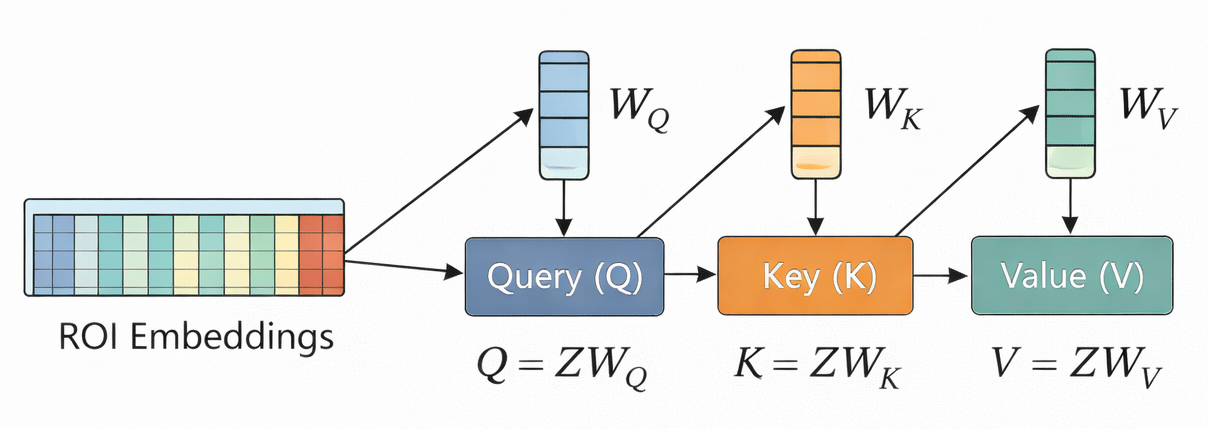}
    \caption{Illustration of the query–key–value (Q–K–V) projection in the self-attention module. ROI-wise embeddings are linearly projected into query, key, and value representations using learnable matrices \(W_Q\), \(W_K\), and \(W_V\), enabling each brain region to attend to functionally relevant regions during attention computation.}
    \label{fig:qkv_attention}
\end{figure}

\begin{figure}[t]
    \centering
    \includegraphics[width=0.78
    \linewidth]{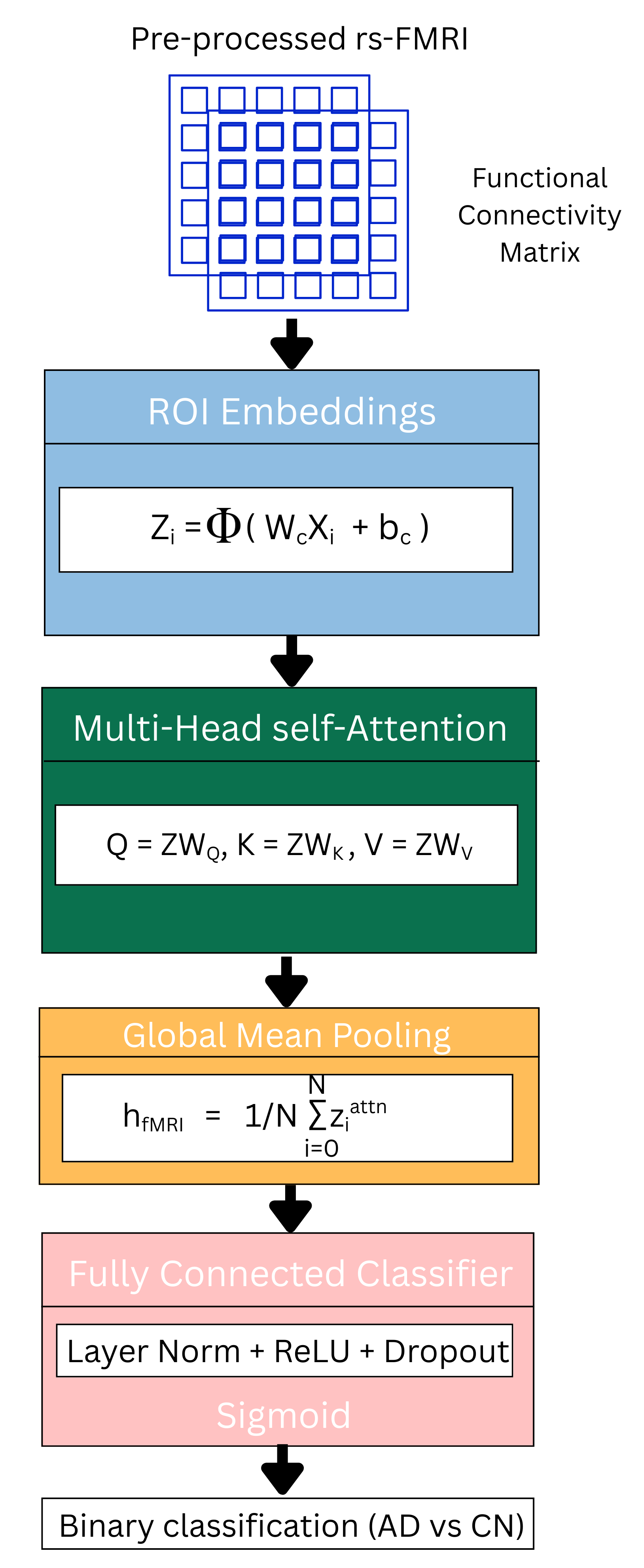}
    \caption{Architecture of the proposed fMRI-only self-attention model. 
Resting-state fMRI is represented as a functional connectivity matrix, 
with each ROI treated as a token and projected into a latent embedding space. 
Multi-head self-attention captures global inter-regional functional interactions, 
followed by global mean pooling and a fully connected classifier to perform 
binary Alzheimer’s disease versus cognitively normal (AD vs CN) classification.}
    \label{fig:fmri_only_architecture}
\end{figure}

The resulting attention-enhanced ROI representations are subsequently aggregated using global mean pooling across the token dimension to obtain a compact visit-level functional representation:
\begin{equation}
\mathbf{h}_{\text{fMRI}} = \frac{1}{N}\sum_{i=1}^{N} \mathbf{z}_i^{\text{attn}},
\end{equation}
where $\mathbf{z}_i^{\text{attn}}$ denotes the output of the self-attention layer for the $i$-th ROI. As illustrated in Figure~\ref{fig:qkv_attention}, the ROI embeddings are linearly projected into query, key, and value representations to compute attention weights that capture inter-regional functional dependencies.An overview of the proposed self-attention–based rs-fMRI classification architecture is illustrated in Figure~\ref{fig:fmri_only_architecture}, with detailed descriptions of each component provided below.

\subsection{Classification Head and Optimization Strategy}
\label{sec:fmri_classification}

The attention-derived visit-level functional representation
$\mathbf{h}_{\text{fMRI}} \in \mathbb{R}^{d}$
is passed to a fully connected classification head to predict the probability of Alzheimer’s disease. To improve training stability and reduce overfitting, the classifier consists of multiple layers with normalization, nonlinear activation, and dropout regularization.

Specifically, the classification head is defined as:
\begin{equation}
\hat{y} = f_{\text{cls}}(\mathbf{h}_{\text{fMRI}}),
\end{equation}
where $f_{\text{cls}}(\cdot)$ denotes a multilayer perceptron composed of layer normalization, linear transformations, rectified linear unit (ReLU) activations, and dropout. The final output $\hat{y} \in \mathbb{R}$ represents a scalar logit corresponding to the Alzheimer’s disease class. During inference, a sigmoid activation is applied to obtain the posterior probability:
\begin{equation}
p(y=1 \mid \mathbf{X}_{\text{fMRI}}) = \sigma(\hat{y}).
\end{equation}

Model parameters are optimized using the AdamW optimizer with a fixed learning rate and weight decay to promote generalization. Training is performed using a class-weighted binary cross-entropy loss to mitigate the effects of class imbalance between cognitively normal (CN) and Alzheimer’s disease (AD) samples. The loss function is defined as:
\begin{equation}
\mathcal{L} =
- w_1 \, y \log \sigma(\hat{y})
- (1 - y) \log \left(1 - \sigma(\hat{y})\right),
\end{equation}
where $w_1$ denotes the positive class weight computed as the ratio of CN to AD samples in the training set.

To prevent information leakage across visits of the same subject, a subject-wise data splitting strategy is employed. Subjects are randomly partitioned into training and test sets using an 80:20 ratio, and all visits belonging to a given subject are assigned exclusively to one split. Results are reported on this single subject-wise split; due to the modest cohort size of 60 subjects, cross-validation across multiple 
folds remains computationally unstable and will be reported in an extended study with a larger cohort.

\section{Experimental Results and Discussion}
\label{ERD}

\subsection{Experimental Setup}
All experiments were conducted using a visit-level evaluation protocol on resting-state fMRI data. The dataset was split into training and test sets using an 80:20 subject-wise partitioning strategy, with all visits of a given subject assigned exclusively to one partition to prevent information leakage across longitudinal visits. The resulting split preserved a balanced representation of AD and CN subjects 
in both sets. Performance metrics are reported on the held-out test set of this single split.The resulting split preserved a balanced representation of Alzheimer’s disease (AD) and cognitively normal (CN) samples in both sets.
\subsection{Hyperparameter Settings}
Table~\ref{tab:hyperparameters} summarizes the hyperparameter configuration used for training the proposed attention-based fMRI model. The network was optimized using the AdamW optimizer with a fixed learning rate, and dropout regularization was applied within both the attention module and the classification head to mitigate overfitting. All hyperparameters were kept constant across experiments.

\begin{table}[h]
\caption{Hyperparameter Settings Used for Model Training}
\label{tab:hyperparameters}
\renewcommand{\arraystretch}{1.05}
\small
\centering
\begin{tabular}{@{}lc@{}}
\toprule
\textbf{Training Parameter} & \textbf{Value} \\
\midrule
Batch size                  & 8 \\
Epochs                      & 40 \\
Learning rate               & $1 \times 10^{-4}$ \\
Optimizer                   & AdamW \\
Loss function               & BCEWithLogitsLoss (class-weighted) \\
Embedding dimension         & 128 \\
Number of attention heads   & 4 \\
Dropout rate                & 0.3 \\
\bottomrule
\end{tabular}
\end{table}

\subsection{Evaluation Metrics}
Model performance was evaluated using accuracy, precision, recall, F1-score, and the area under the receiver operating characteristic curve (ROC-AUC). Accuracy reflects overall classification correctness, while precision and recall quantify the reliability and sensitivity of Alzheimer’s disease detection, respectively. The F1-score provides a balanced measure between precision and recall, and ROC-AUC evaluates the model’s threshold-independent discriminative ability, which is particularly important for imbalanced medical datasets.

\subsection{Results and Discussion}

The classification performance of the proposed attention-based rs-fMRI model is summarized in Table~\ref{tab:results}.

\begin{table}[h]
\caption{Performance of the Attention-Based rs-fMRI Model}
\label{tab:results}
\renewcommand{\arraystretch}{1.0}
\small
\centering
\begin{tabular}{@{}lc@{}}
\toprule
\textbf{Metric} & \textbf{Value} \\
\midrule
Accuracy        & 0.8895 \\
ROC-AUC         & 0.9000 \\
Precision       & 0.9000 \\
Recall          & 0.8182 \\
F1-score        & 0.8571 \\
\bottomrule
\end{tabular}
\vspace{2pt}
\end{table}

The proposed attention-based rs-fMRI model achieves strong and balanced classification performance, demonstrating its effectiveness in capturing disease-related functional connectivity alterations associated with Alzheimer's disease. An overall accuracy of 88.95\% and a ROC-AUC of 0.90 indicate robust discriminative capability between AD 
and cognitively normal subjects. The high precision of 0.90 suggests that the model reliably identifies Alzheimer's disease cases with a low false-positive rate, which is particularly desirable in clinical screening scenarios. A recall of 0.8182 reflects effective sensitivity to disease-related functional disruptions, demonstrating the model's ability to detect AD cases despite the inherent variability and noise present in resting-state fMRI data. The F1-score of 0.8571 highlights a favorable balance between precision and recall, confirming that the attention mechanism successfully emphasizes disease-relevant functional connections while suppressing less informative interactions.

Prior deep learning studies on rs-fMRI--based Alzheimer's disease classification have reported varying performance depending on cohort size, input representation, and evaluation protocol. Methods operating on raw 4D spatiotemporal fMRI data or large multi-site cohorts have 
reported higher accuracies~\cite{LI2020280}, while approaches constrained to modest longitudinal cohorts with functional connectivity--based representations face inherently greater generalization challenges due to limited training data and inter-subject variability. The proposed attention-based framework achieves an accuracy of 88.95\% 
and ROC-AUC of 0.90 on a 60-subject longitudinal ADNI cohort using only FC matrices, demonstrating that self-attention over functional connectivity provides an effective and interpretable alternative to volumetric or spatiotemporal approaches, particularly in data-constrained clinical settings. Direct numerical comparison with existing methods is precluded by differences in cohort composition, input modality, task formulation, and evaluation protocol ~\cite{LI2020280, kan2022brain}; the proposed method is nonetheless notable for its interpretable attention-based connectivity modeling 
on a modestly sized longitudinal cohort without reliance on raw volumetric or temporal fMRI representations.


\section{Conclusion}\label{con}
This study presents an attention-based deep learning framework for Alzheimer’s disease classification using resting-state fMRI functional connectivity data from the ADNI cohort. By treating brain regions as tokens and applying self-attention mechanisms to functional connectivity matrices, the proposed approach effectively captures global and long-range interactions among distributed brain networks that are disrupted in Alzheimer’s disease.

Experimental results demonstrate that the proposed rs-fMRI attention model achieves strong and balanced classification performance, with a high ROC-AUC and favorable precision–recall trade-offs under a subject-wise longitudinal evaluation protocol. These findings indicate that self-attention provides an effective mechanism for emphasizing disease-relevant functional connections while mitigating the effects of noise and inter-subject variability inherent in resting-state fMRI data. The use of class-weighted optimization further contributes to stable learning in the presence of mild class imbalance, reflecting realistic clinical data distributions.

Overall, this work highlights the effectiveness of attention-driven modeling of resting-state functional connectivity for neuroimaging-based Alzheimer’s disease detection. The proposed framework is interpretable, scalable, and well suited for longitudinal analysis. In future work, this approach will be extended to incorporate multimodal fusion of structural MRI and rs-fMRI data using attention-based integration strategies, with the goal of jointly modeling anatomical degeneration and functional network disruption to further improve diagnostic performance and clinical interpretability.


\bibliographystyle{ieeetr}
\bibliography{Stats}

\end{document}